\documentclass[onecolumn,12pt]{IEEEtran}
\ifCLASSINFOpdf
\else
\fi
\usepackage{cite}
\usepackage{setspace}
\usepackage{amsfonts}
\usepackage{amssymb}
\usepackage{graphicx}
\usepackage[cmex10]{amsmath}
\usepackage{algorithm}
\usepackage{algorithmic}
\usepackage{multirow}
\usepackage{bm}
\usepackage{booktabs}
\usepackage{graphics}
\usepackage{graphicx}
\usepackage{epsfig}

\newtheorem{remark}{Remark}
\newtheorem{proposition}{Proposition}

\usepackage{fancyhdr}

\makeatletter
\newcommand\figcaption{\def\@captype{figure}\caption}
\makeatother

\IEEEoverridecommandlockouts
\begin{document}
\title{Backhaul-Aware User Association and Resource Allocation for Energy-Constrained HetNets}
\author{%
Qiaoni Han,~Bo~Yang,\emph{~Member,~IEEE},~Guowang~Miao,\emph{~Member,~IEEE},\\
~Cailian~Chen,\emph{~Member,~IEEE}, Xiaocheng Wang, and~Xinping~Guan,\emph{~Senior~Member,~IEEE}
\thanks{Copyright (c) 2015 IEEE. Personal use of this material is permitted. However, permission to use this material for any other purposes must be obtained from the IEEE by sending a request to pubs-permissions@ieee.org.}
\thanks{Q. Han, B. Yang, C. Chen, and X. Guan are with the Department of Automation, Shanghai Jiao Tong University, Shanghai 200240, China, and also with the Key Laboratory of System Control and Information Processing, Ministry of Education of China, Shanghai 200240, China (e-mail: qiaoni@sjtu.edu.cn; bo.yang@sjtu.edu.cn; cailianchen@sjtu.edu.cn; xpguan@sjtu.edu.cn). (Corresponding author: Bo Yang.)}
\thanks{G. Miao is with KTH Royal Institute of Technology, Stockholm 10044, Sweden (e-mail: guowang@kth.se).}
\thanks{X. Wang is with the Department of Electronic Engineering, Shanghai Jiao Tong University, Shanghai 200240, China (e-mail: wangxiaocheng@sjtu.edu.cn).}}
\vspace{-1.5cm}
\maketitle
\thispagestyle{fancy}
\fancyhead[L]{DOI: 10.1109/TVT.2016.2533559}
\cfoot{}
\vspace{-0.8cm}
\begin{spacing}{2.0}
\begin{abstract}
Growing attentions have been paid to renewable energy or hybrid energy powered heterogeneous networks (HetNets). In this paper, focusing on backhaul-aware joint user association and resource allocation for this type of HetNets, we formulate an online optimization problem to maximize the network utility reflecting proportional fairness. Since user association and resource allocation are tightly coupled not only on resource consumption of the base stations (BSs), but also in the constraints of their available energy and backhaul, the closed-form solution is quite difficult to obtain. Thus, we solve the problem distributively via employing some decomposition methods. Specifically, at first, by adopting primal decomposition method, we decompose the original problem into a lower-level resource allocation problem for each BS, and a higher-level user association problem. For the optimal resource allocation, we prove that a BS either assigns equal normalized resources or provides equal long-term service rate to its served users. Then, the user association problem is solved by Lagrange dual decomposition method, and a completely distributed algorithm is developed. Moreover, applying results of the subgradient method, we demonstrate the convergence of the proposed distributed algorithm. Furthermore, in order to efficiently and reliably apply the proposed algorithm to the future wireless networks with an extremely dense BS deployment, we design a virtual user association and resource allocation scheme based on the software-defined networking architecture. Lastly, numerical results validate the convergence of the proposed algorithm and the significant improvement on network utility, load balancing and user fairness.
\end{abstract}

\ifCLASSOPTIONpeerreview
\begin{center} \bfseries EDICS Category: 3-BBND \end{center}
\fi

\IEEEpeerreviewmaketitle
\section{Introduction}
\label{intro}
\subsection{Motivation}
Nowadays, with the proliferation of smartphones, tablets, video streaming, and emerging new applications and services, the data traffic demand in cellular networks grows tremendously \cite{1,vf1}. To meet this demand, cellular networks are trending strongly towards increasing heterogeneity, especially through overlapping deployment of small cell base stations (BSs), e.g., microcells, picocells and femtocells, which differ primarily in terms of maximum transmit power, physical size, ease-of-deployment and cost \cite{2,3,4,vf2,v1}. In fact, it has been widely accepted that heterogeneous network (HetNet) is a promising approach to achieve high spectral and energy efficiency. On the other hand, the unprecedented data traffic growth and rapidly increasing deployment of small cells have pushed the limits of energy consumption in wireless networks, which causes both serious environment problem and sharp rising energy cost for network operators \cite{1,v2,v3,vf3,5}. Then, the economic and ecological concerns, together with the advance of energy harvesting technologies, have advocated the ``green communications" solutions, where cellular BSs, especially small cell BSs, are powered by renewable energy sources (RES) such as sustainable biofuels, solar and wind energy \cite{5,6,vf4,7}. Moreover, due to long-term cost savings, easier deployment and reduced carbon emissions, HetNets with renewable energy powered BSs can be a sustainable and economically convenient solution for next-generation cellular networks.

As cellular networks evolve into dense, organic and irregular HetNets, load awareness has been elevated to a central problem \cite{8}. Primarily, in HetNets, due to the disparities of transmit powers and BS capabilities, even if users are uniformly distributed in geography, the well-known user association schemes based on ``natural" metrics like signal-to-interference-plus-noise ratio (SINR) or received signal strength indicator (RSSI) can lead to an extreme load imbalance among macro BSs and small cell BSs \cite{9,v4}. Furthermore, the key performance metric for a user should be the rate, not SINR. The rate is of course directly related to SINR (e.g., $\log_2(1+\text{SINR})$), but it is also proportional to the fraction of resources that the user gets. Thus, heavily-loaded cells may provide lower rate over time, even though they offer a higher SINR \cite{8}. As a consequence, a joint user association and resource allocation strategy that is able to both introduce load balancing and improve network-wide performance is needed. While for HetNets with emerging RES, owning to the dynamics of renewable energy generation and limited capacity of energy storage, sole RES may not guarantee enough power supplies for all kinds of BSs. Hence, future HetNets are more likely to adopt hybrid energy supplies with both traditional electric grid and RES \cite{10}. In this case, since the available energy of RES varies with time and space and the average electric power consumed by BSs should be limited to save the cost, HetNets call for load balancing strategy that takes into account the constrained energy supplies. Moreover, the large number of small cell BSs to be deployed in HetNets may incur overwhelming traffic over backhaul links. However, the current small cell backhaul solutions, such as xDSL, non-line-of-sight microwave, are far from the ideal ones providing sufficiently large data rate \cite{11,12}. As such, the backhaul (data rate) constraint has become increasingly stringent in HetNets.
\subsection{Related Works}
So far, a lot of works on load balancing for HetNets have been presented. Andrews \emph{et al.} in [8] survey the technical issues and primary approaches on load balancing in HetNets. The works in \cite{13,14,15,16,17} deal with joint user association and resource allocation for HetNets powered by grid energy. When it comes to HetNets with RES, some studies on user association and/or resource allocation have been presented \cite{18,vf5,19,20,21,22}. Specifically, for HetNets solely powered by RES, \cite{18} proposes load balancing scheme among different BSs in order to obtain higher downlink network utility, and \cite{vf5} designs offline and online load balancing schemes, which are both energy-aware and QoS-aware. Then, for hybrid energy powered HetNets, to reduce on-grid energy consumption by maximizing the utilization of green energy, \cite{19} presents a distributed scheme to enable green-energy aware and latency aware (GALA) user-BS associations, and \cite{20} formulates and solves a joint multi-stage energy allocation and multi-BS energy balancing problem. On the other hand, to strive for a balance between the average traffic delivery latency and green energy utilization, \cite{21} proposes a virtually distributed algorithm, and \cite{22} develops a network utility aware (NUA) traffic load balancing scheme.

Different from the studies only on user association in \cite{19,20,21,22} and the references therein \cite{8}, this paper focuses on joint user association and resource allocation for HetNets. For easy implementation and catering to the existing standard model for LTE, unique association of users, which means each user can be associated to at most one BS at a time, is considered. This is different from the works in \cite{16} and \cite{17}, where users can be served by multiple BSs simultaneously. Moreover, we take both backhaul and energy constraints into account, which is distinguished from the works in \cite{13,14,15,18,19,20,21}. Then, the formulated problem has not only mixed integer variables but also coupled variables in the constraints of resources, energy and backhaul. Employing some decomposition methods, we efficiently obtain the condition for two kinds of resource partition of each BS and develop a completely distributed algorithm for user association, which are the novelty and main contributions of this paper.
\subsection{Contributions}
We deal with the backhaul-aware joint user association and resource allocation problem for hybrid energy powered HetNets, which have both backhaul and energy constraints. The contributions of this paper are summarized below.
\begin{itemize}
\item We consider a hybrid energy powered HetNet with limited backhaul and take into account the unique association of users, which can be easily extended to general energy-constrained LTE HetNets with wired or wireless backhaul connections. With the objective of maximizing network utility reflecting proportional fairness (PF), we formulate an online optimization problem. However, its closed-form solution is quite difficult to obtain in threefold: both binary and continuous variables; tightly coupled variables in multiple constraints; and desired distributed solutions for each user and BS.
\item Due to the coupling and non-convexity, we employ several decomposition methods to solve the formulated problem. Firstly, by applying primal decomposition method, we decompose the original problem into a lower-level resource allocation problem and a higher-level user association problem. Secondly, we further decompose the resource allocation problem into subproblems for each BS. Then, based on complementary slackness property, we efficiently obtain a BS's optimal resource partition, which either assigns equal normalized resources or provides equal long-term service rate to its served users. Thirdly, given optimal resource partition of BSs, we solve the user association problem with Lagrange dual decomposition method, and develop a completely distributed algorithm for joint user association and resource allocation.
\item For the efficient and reliable application of the proposed algorithm to the future wireless networks with an extremely dense BS deployment, we design a virtual user association and resource allocation (vUARA) scheme based on software-defined networking (SDN) architecture. Since virtual users and virtual BSs (vBSs) can be generated in the radio access networks controller (RANC) to emulate a distributed user association and resource allocation solution that requires interactive adjustments between users and BSs, the proposed scheme significantly reduces the communication overhead over air interface and avoids the information leaking to users.
\item We demonstrate the convergence of the proposed distributed algorithm both by applying results of the subgradient method in theory and by numerical results with random network realizations. Moreover, compared with max-SINR association and range expansion association, numerical results indicate that, the proposed algorithm provides a significant improvement on network utility, load balancing and user fairness. Furthermore, we present and analyze the effects of energy and backhaul constraints on algorithm performance.
\end{itemize}

The rest of the paper is organized as follows. We describe the energy-constrained HetNet under consideration in Section \ref{sys mod}, and formulate the optimization problem in Section \ref{pro formu}. The solution to the formulated problem is presented in Section \ref{solu}. Section \ref{imple} gives implementation of the vUARA scheme. Simulation studies are provided in Section \ref{per eva}, and Section \ref{conclu} concludes the paper.

\section{System Model}
\label{sys mod}
\subsection{An Energy-Constrained HetNet}
\begin{figure}[htbp]
  \centering
  \includegraphics[width=11.0cm]{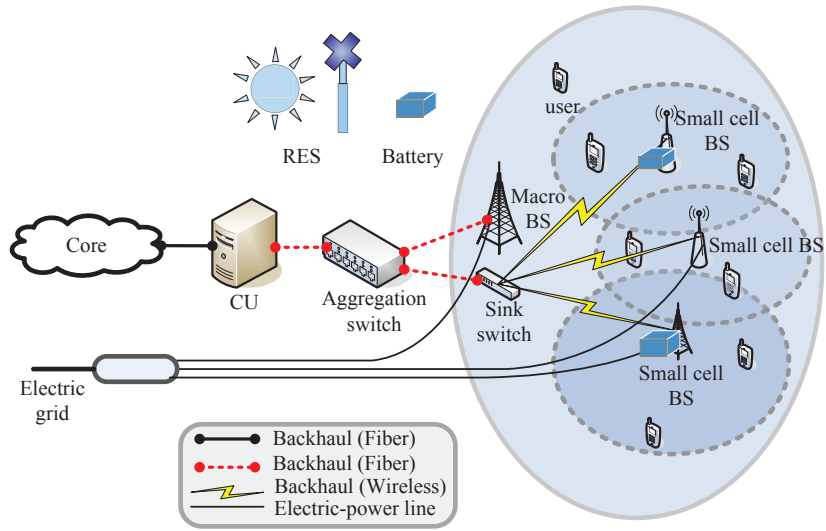}\\
  \caption{Network model}\label{fig:1}
\end{figure}

We consider a HetNet composed of $A$-tiers of BSs, where each tier models a particular type of BSs. For example, let $A=3$, then tier 1 consists of traditional macro BSs, tier 2 and tier 3 may be comprised of pico BSs and femto BSs, respectively. Generally, macro BSs, pico BSs and femto BSs have different transmission power and coverage. Moreover, for BSs of different tiers, their energy consumption and backhaul capacities vary widely, and the coefficients of path loss between them and the users are also different. All the BSs are assumed to be connected by a high speed backhaul through which information exchange with negligible delay is possible. Furthermore, given such a HetNet, we adopt a hybrid energy supply, i.e., the BSs are powered by traditional electric grid, RES, or both. For BSs with RES, the RES collects energy from the environment by solar panels and/or windmills and charges the corresponding battery (each BS has its own RES and battery, which vary with the size of the BS). This is especially important in rural areas, where the access to electric grid may be impossible or too expensive. A network model for such HetNet with backhaul layout is given in Fig. \ref{fig:1}, where pico BSs and femto BSs are all treated as small cell BSs.

We denote by $\mathcal{N}=\{1,2,\cdots,N\}$ and $\mathcal{K}=\{1,2,\cdots,K\}$ the sets of all BSs and all single antenna users, respectively. Each BS in set $\mathcal{N}$ schedules transmissions over contiguous time-frequency slots, referred to as RBs, each comprising a block of OFDM subcarriers and symbols. Moreover, the transmitting power of BS $n\in\mathcal{N}$ on each RB is assumed to be always constant.

\subsection{Transmission Model}
According to \cite{20}, the duration of one association period, which refers to a time duration in which the BSs and the users involved have stable associations, depends on the dynamics of renewable energy generation and mobile traffic intensity. Here, we omit the specific definition of these periods, and simply denote the association period by index $l\in\mathcal{L}=\{1,\cdots,L\}$, and its time duration as $s^{(l)}$. In following subsections, we focus on the $l$th association period.

At time instant $t$ of the $l$th association period, we denote the channel power gain between user $k\in\mathcal{K}$ and BS $n\in\mathcal{N}$ by $H_{nk}^{(l)}(t)$ as
\begin{equation}
H_{nk}^{(l)}(t)=G_{nk}^{(l)}(t)F_{nk}^{(l)}(t),\forall (n,k)\in \mathcal{N}\times\mathcal{K}, \label{t1}
\end{equation}
where $G_{nk}^{(l)}(t)$ is the large-scale slow fading component capturing effects of path-loss and shadowing, and $F_{nk}^{(l)}(t)$ represents the small-scale fast fading component. ``$\cdot\times\cdot$" denotes the Cartesian product of two sets. Moreover, it is assumed that $G_{nk}^{(l)}(t)$ remains constant during one association period. Thus, it can be simplified as $G_{nk}^{(l)}$. To model the small-scale Rayleigh fading, it is assumed that $F_{nk}^{(l)}(t)$ fluctuates fast during an association period following an exponential probability distribution function with variance 1 \cite{14}.

The instantaneous SINR from BS $n$ to user $k$ on a RB is
\begin{equation}
\text{SINR}_{nk}^{(l)}(t)=\frac{P_n^{(l)}H_{nk}^{(l)}(t)}{\sum\nolimits_{j\not=n}P_j^{(l)}H_{jk}^{(l)}(t)+\sigma^2}, \label{t2}
\end{equation}
where $P_n^{(l)}$ and $P_j^{(l)}$ are transmission power of BS $n$ and $j$ on a RB in the association period $l$, respectively, and $\sigma^2$ denotes the thermal noise spectral power.

Accordingly, the instantaneous achievable rate at user $k$, if it is served by BS $n$, can be written as
\begin{equation}
R_{nk}^{(l)}(t)=B_0\log_2(1+\text{SINR}_{nk}^{(l)}(t)), \label{t3}
\end{equation}
where $B_0$ denotes the bandwidth over which a RB is realized.

In general, since the mobility of users between BSs takes place at larger time scales and channel may vary during the whole association period, the association decision should be taken considering the long-term rate that a given user will obtain if it is associated with a certain BS. Referring to \cite{14}, the long-term SINR from BS $n$ to user $k$ on a RB can be written as
\begin{equation}
\overline{\text{SINR}}_{nk}^{(l)}=\frac{P_n^{(l)}G_{nk}^{(l)}}{\sum\nolimits_{j\not=n}P_j^{(l)}G_{jk}^{(l)}+\sigma^2}, \label{t4}
\end{equation}
and the corresponding long-term rate is
\begin{equation}
R_{nk}^{(l)}=B_0\log_2\left(1+\overline{\text{SINR}}_{nk}^{(l)}\right). \label{t5}
\end{equation}

On the other hand, the number of users associated with a BS is usually more than one, and the users of the same BS need to share the resources. Thus, the long-term service rate achieved by a user depends on the load of the BS and will only be a fraction of the rate $R_{nk}^{(l)}$ (unless BS $n$ exclusively serves user $k$ in association period $l$). That is to say, a user's long-term service rate for its association with a BS depends on both the load and the resource partition method of the BS. In this paper, we assume unique association of users and define the user association indicator $x_{nk}^{(l)}$ as
\begin{equation}
x_{nk}^{(l)}=
\begin{cases}
1,\,\,\text{if user $k$ is associated with BS $n$,}\\
0,\,\,\text{otherwise.} \label{t6}
\end{cases}
\end{equation}

In addition, denoting the fraction of RBs over which user $k$ is served by BS $n$ as $y_{nk}^{(l)}$, the overall rate of user $k$ and the required backhaul of BS $n$ can be respectively given as
\begin{equation}
R_k^{(l)}=\sum_{n\in \mathcal{N}}x_{nk}^{(l)}y_{nk}^{(l)}R_{nk}^{(l)}W_n, \label{t7}
\end{equation}
\begin{equation}
Z_n^{(l)}=\sum_{k\in \mathcal{K}}x_{nk}^{(l)}y_{nk}^{(l)}R_{nk}^{(l)}W_n, \label{t8}
\end{equation}
where $W_n$ is the total number of RBs of BS $n$, and may vary for different BSs.

Since the backhaul of BSs is considered to be constrained, for BS $n\in\mathcal{N}$, there is
\begin{equation}
Z_n^{(l)}\leq Z_{n,\text{bh}}, \label{t9}
\end{equation}
where $Z_{n,\text{bh}}$ is BS $n$'s maximum backhaul capacity that differs considerably among the $A$-tiers of BSs.

\subsection{BSs' Energy Consumption and Constraints}
Firstly, adopting the model in EARTH project \cite{23}, the energy consumption of BS $n$ in association period $l$ can be expressed as
\begin{equation}
C_n^{(l)}=s^{(l)}\left[P_{n0}+\delta_n\left(\sum\nolimits_{k\in \mathcal{K}}x_{nk}^{(l)}y_{nk}^{(l)}W_n\right)P_n^{(l)}\right], \label{t10}
\end{equation}
where $P_{n0}$ is the fixed part of BS $n$'s power consumption, $\delta_n$ is the variable power consumption slope, and $\delta_n\left(\sum\nolimits_{k\in \mathcal{K}}x_{nk}^{(l)}y_{nk}^{(l)}W_n\right)P_n^{(l)}$ is the variable part related to the transmission power and the consumed resources. It should be noted that, for BSs belonging to different tiers, their fixed power consumption, power consumption slopes and transmission power vary widely \cite{3,23}.

Then, due to RES and batteries adopted by some BSs, we analyze their battery dynamics.

In order to provide a general model for energy harvesting BSs, we do not assume a particular type of energy harvester. Specifically, we assume that BS $n$ is powered by both RES and electric grid. Let $B_n^{(l)}$ and $E_n^{(l)}$ respectively be the battery level and the renewable energy arriving at the beginning of association period $l$. Here, $E_n^{(l)}$ can also be viewed as the energy arriving in the association period $l-1$, but is not used until period $l$. This assumption is reasonable since, in practice, user association and resource allocation decisions can only be made based on the battery state information available at the beginning of the association period. Besides, it is assumed that, for a larger BS that is powered by RES, the corresponding RES power and battery capacity are also larger.

At the beginning of association period $l+1$, the battery level of BS $n$ is updated as
\begin{equation}
B_n^{(l+1)}=f\left(B_n^{(l)},C_{n,\text{E}}^{(l)},E_n^{(l+1)}\right), \label{t11}
\end{equation}
where $f(\cdot)$ depends upon the battery dynamics, such as storage efficiency and memory effects \cite{22}. $C_{n,\text{E}}^{(l)}$ is BS $n$'s energy consumption from RES in association period $l$. A common practice is to consider the battery update as
\begin{equation}
B_n^{(l+1)}=\max\left\{0,\min\left\{B_n^{(l)}-C_{n,\text{E}}^{(l)}+E_n^{(l+1)},B_{n\text{max}}\right\}\right\}, \label{t12}
\end{equation}
where $B_{n\text{max}}$ denotes BS $n$'s maximum battery capacity, the inner minimization accounts for possible battery overflows, and the outer maximization assures the non-negativity of the battery levels. In general, $C_{n,\text{E}}^{(l)}$ will be limited by a function of the current battery level, i.e.,
\begin{equation}
C_{n,\text{E}}^{(l)}\leq g_n^{(l)}(B_n^{(l)}), \label{t13}
\end{equation}
where $g_n^{(l)}(\cdot)$ limits the renewable energy that can be used in association period $l$ in order to spend the energy in a more conservative way. Simply, we can consider that only a given fraction of the battery is allowed to be used.

At the same time, in order to save the cost on electric power, we give BS $n$'s limited average electric power, i.e., the electric grid energy can be used freely on the condition that the average power over all association periods is no greater than the threshold $G_{n,\text{ave}}$ \cite{24},
\begin{equation}
\frac{1}{L}\sum_{l\in \mathcal{L}}C_{n,\text{G}}^{(l)}=\frac{1}{L}\sum_{l\in \mathcal{L}}\left[C_n^{(l)}-C_{n,\text{E}}^{(l)}\right]\leq G_{n,\text{ave}}, \label{t14}
\end{equation}
where $C_{n,\text{E}}^{(l)}=0$, if BS $n$ is solely powered by the electric grid.

Transforming the average electric power into each association period and combining (\ref{t13}), we have
\begin{equation}
C_n^{(l)}=C_{n,\text{E}}^{(l)}+C_{n,\text{G}}^{(l)}\leq g_n^{(l)}(B_n^{(l)})+G_{n,\text{ave}}, \label{t15}
\end{equation}
which can be regard as the energy constraint for BS $n$ in association period $l$.

\begin{remark}
\label{rmk1} In the above description, we didn't give any notation indicating the specific energy supplies of BSs. This is due to the fact that, for BSs without RES, we can set $g_n^{(l)}(B_n^{(l)})=0$; while for BSs without energy supplies from electric grid, we have $G_{n,\text{ave}}=0$.
\end{remark}

\section{Problem Formulation}
\label{pro formu}
In practice, only causal information, i.e., information of the past and current channel states and energy harvesting, is available. Thus, an online approach is quite desirable. In addition, based on the analysis of BSs' energy consumption and constraints, we can address the user association and resource allocation problem on each association period. In the following sections, the superscript $(l)$ is omitted.

Firstly, taking a utility function perspective, we assume that user $k$ obtains utility $U_k(R_k)$. In order to achieve a desired balance between network-wide performance and user fairness, we shall choose a continuously differentiable, monotonically increasing, and strictly concave utility function. Such that, larger rate yields greater utility, and the shape of the concave function also imposes some desired notion of fairness. Here, we consider the well-known PF, which is imposed by choosing the utility function
\begin{equation}
U_k\left(R_k\right)=\log \left(R_k\right). \label{t16}
\end{equation}

Then, the network utility maximization problem is given as:
\begin{align}
\textbf{P1}:\max\limits_{\bm{x},\bm{y}}\,&\sum_{k\in\mathcal{K}}\log \left(\sum_{n\in\mathcal{N}}x_{nk}y_{nk}R_{nk}W_n\right) \label{t17}\\
  \mbox{s.t.}\,\,&\sum_{n\in\mathcal{N}}x_{nk}=1,\forall k\in\mathcal{K} \tag{17a}\\
  &\sum_{k\in\mathcal{K}}x_{nk}y_{nk}\leq 1,\forall n\in\mathcal{N} \tag{17b}\\
  &\sum_{k\in \mathcal{K}}x_{nk}y_{nk}R_{nk}W_n\leq Z_{n,\text{bh}},\forall n\in\mathcal{N}  \tag{17c}\\
  &C_n\leq g_n(B_n)+G_{n,\text{ave}},\forall n\in\mathcal{N} \tag{17d}\\
  &x_{nk}\in \{0,1\},\forall n\in\mathcal{N},\forall k\in\mathcal{K} \tag{17e}\\
  &y_{nk}\in[0,1],\forall n\in\mathcal{N},\forall k\in\mathcal{K} \tag{17f}
\end{align}
where (17a) denotes that each user can only be associated with one BS at a given time; (17b), (17c) and (17d) respectively specify the constraints of resources, backhaul and energy for all the BSs; (17e) and (17f) keep the association indicators binary and the resource allocation variables being between 0 and 1, respectively.

Since $x_{nk}$s take binary values and $\sum_{n\in\mathcal{N}}x_{nk}=1$, we have
\begin{equation}
\begin{split}
&\sum_{k\in\mathcal{N}}\log \left(\sum_{n\in\mathcal{N}}x_{nk}y_{nk}R_{nk}W_n\right)\\
&=\sum_{k\in\mathcal{K}}\sum_{n\in\mathcal{N}}x_{nk}\log \left(y_{nk}R_{nk}W_n\right).  \label{t18}
\end{split}
\end{equation}

Unfortunately, due to binary variables $x_{nk}$s and continuous variables $y_{nk}$s, \textbf{P1} is a mixed integer nonlinear programming (MINLP) problem, which is generally NP-hard. Moreover, the variables $x_{nk}$s and $y_{nk}$s are coupled in the constraints of resources, backhaul and energy, which brings much more complexity. Furthermore, the optimal solution should be distributed for each user and BS. In the following section, we focus on solving \textbf{P1} distributively by employing some decomposition methods.

\begin{remark}
\label{rmk2} The above joint user association and resource allocation problem is a MINLP problem. MINLP problems have the difficulties of both of their sub-classes, i.e., the combinatorial nature of mixed integer programming (MIP) and the difficulty in solving nonlinear programming (NLP). Since both MIP and NLP are NP-complete, the joint user association and resource allocation problem \textbf{P1} is NP-hard \cite{minlp}.
\end{remark}

\section{Solution}
\label{solu}
In this section, the solution to the formulated problem is detailed, and a completely distributed algorithm is developed for the backhaul-aware joint user association and resource allocation in energy-constrained HetNets.

\subsection{Primal Decomposition}
Obviously, the choices of $y_{nk}$s rely on the values of $x_{nk}s$. Given these coupled variables, we can apply the primal decomposition method to decompose \textbf{P1} into the following problems in two levels. Firstly, by fixing variables $x_{nk}$s, we have the lower-level problem:
\begin{align}
  \textbf{P2}:\max\limits_{\bm{y}}\,&\sum_{n\in\mathcal{N}}\sum_{k\in\mathcal{K}_n}\log \left(y_{nk}R_{nk}W_n\right) \label{t19}\\
  \mbox{s.t.}\,\,&\sum_{k\in\mathcal{K}_n}y_{nk}\leq 1,\forall n\in\mathcal{N} \tag{19a}\\
  &\sum_{k\in\mathcal{K}_n}y_{nk}R_{nk}W_n\leq Z_{n,\text{bh}},\forall n\in\mathcal{N} \tag{19b}\\
  &\sum_{k\in\mathcal{K}_n}y_{nk}\leq \frac{\left[g_n(B_n)+G_{n,\text{ave}}\right]/s-P_{n0}}{\delta_nW_nP_n},\forall n\in\mathcal{N} \tag{19c}\\
  &0\leq y_{nk}\leq 1,\forall k\in\mathcal{K}_n,\forall n\in\mathcal{N} \tag{19d}
\end{align}
where $\mathcal{K}_n$ denotes the set of users that associated with BS $n$, and $|\mathcal{K}_n|=\sum_{k\in\mathcal{K}}x_{nk}$.

Then, when $y_{nk}$s are fixed, the higher-level problem (or the master problem) is given by
\begin{align}
  \textbf{P3}:\max\limits_{\bm{x}}\,&\sum_{k\in\mathcal{K}}\sum_{n\in\mathcal{N}}x_{nk}\log \left(y_{nk}R_{nk}W_n\right) \label{t20}\\
  \mbox{s.t.}\,\,&\sum_{n\in\mathcal{N}}x_{nk}=1,\forall k\in\mathcal{K} \tag{20a}\\
  &x_{nk}\in \{0,1\},\forall n\in\mathcal{N},\forall k\in\mathcal{K} \tag{20b}
\end{align}

Since there are no couplings among the subproblems, \textbf{P2} can be further decomposed into $N$ subproblems for all the BSs:
\begin{align}
  \textbf{P4}:\max\limits_{\bm{y}}\, &\sum_{k\in\mathcal{K}_n}\log \left(y_{nk}R_{nk}W_n\right) \label{t21}\\
  \mbox{s.t.}\, &\sum_{k\in\mathcal{K}_n}y_{nk}\leq Q_{n,\text{P}} \tag{21a}\\
   &\sum_{k\in\mathcal{K}_n}y_{nk}R_{nk}W_n\leq Z_{n,\text{bh}} \tag{21b}\\
   &0\leq y_{nk}\leq 1,\forall k\in\mathcal{K}_n \tag{21c}
\end{align}
where
\begin{equation}
Q_{n,\text{P}}=\min\left\{1,\frac{\left[g_n(B_n)+G_{n,\text{ave}}\right]/s-P_{n0}}{\delta_nW_nP_n}\right\}  \label{t22}
\end{equation}
is obtained based on (19a) and (19c), and can be regarded as the normalized resources that BS $n$ can afford with constraints from both overall RBs and available energy.
\vspace{-0.2cm}
\subsection{Resource Allocation with Fixed User Association}
Defining Lagrange multipliers $\mu_n$ and $\nu_n$, the Lagrangian function of \textbf{P4} is given as
\begin{equation}
\begin{split}
L=&\sum_{k\in\mathcal{K}_n}\log \left(y_{nk}R_{nk}W_n\right)-\mu_n\left(\sum_{k\in\mathcal{K}_n}y_{nk}-Q_{n,\text{P}}\right)\\
&- \nu_n\left(\sum_{k\in\mathcal{K}_n}y_{nk}R_{nk}W_n-Z_{n,\text{bh}}\right). \label{t23}
\end{split}
\end{equation}

Applying Karush-Kuhn-Tucker (KKT) conditions, we obtain
\begin{equation}
y_{nk}^*=\frac{1}{\mu_n^*+\nu_n^*R_{nk}W_n},\forall k\in\mathcal{K}_n.  \label{t24}
\end{equation}

It is easily observed that, $y_{nk}$ is completely dependent on dual variables $\mu_n$ and $\nu_n$. Intuitively, $\mu_n$ and $\nu_n$ can be interpreted as the prices of BS $n$ determined by load situation and backhaul state, respectively. When BS $n$ is overloaded (i.e., $\sum_{k\in\mathcal{K}_n}y_{nk}\geq Q_{n,\text{P}}$) or its backhaul is overflowed (i.e., $\sum_{k\in\mathcal{K}_n}y_{nk}R_{nk}W_n\geq Z_{n,\text{bh}}$), the corresponding price $\mu_n$ or $\nu_n$ goes up, then a user $k$'s resource fraction $y_{nk}$ from associating with it decreases. Otherwise, the prices go down, $y_{nk}$ increases, and BS $n$ becomes more attractive.

Generally, this KKT system can be solved by finding the appropriate dual variables $(\mu_n,\nu_n)$. To achieve this, a two-dimensional search or classic constrained optimization techniques such as subgradient method or augmented Lagrangian can be applied. However, both of them are computationally intensive and might not be practical for large-scale problem. Therefore, in the following, we focus on solving this KKT system efficiently based on its specific construction.

On one hand, we note that both the constraints (21a) and (21b) in \textbf{P4} are monotonic functions of $y_{nk}$, and $y_{nk}$ is also a monotonic function of $\mu_n$ when $\nu_n$ is fixed and vice versa. On the other hand, the complementary slackness in constrained optimization states that, for inequality constraints $f_i(\bm x)\leq 0$ that are tight with equality, the associated dual variables are non-zero. Using this result, at a local optimum, each BS can be either energy-constrained or backhaul-constrained. Thus, we can perform our search on two single dual variables instead of a two-dimensional search.

Specifically, if BS $n$ is energy-constrained, i.e.,
\begin{equation}
\begin{cases}
\sum_{k\in\mathcal{K}_n}y_{nk}^*=Q_{n,\text{P}}, \\
\sum_{k\in\mathcal{K}_n}y_{nk}^*R_{nk}W_n\leq Z_{n,\text{bh}},
\end{cases} \label{t25}
\end{equation}
then, there is
\begin{equation}
\begin{cases}
\mu_n^*\ge 0,\\
\nu_n^*=0.
\end{cases} \label{t26}
\end{equation}

Substituting (\ref{t24}) and (\ref{t26}) into the first equation of (\ref{t25}), we obtain
\begin{equation}
y_{nk}^*=\frac{1}{\mu_n^*}=\frac{Q_{n,\text{P}}}{|\mathcal{K}_n|},\forall k\in\mathcal{K}_n, \label{t27}
\end{equation}
which means that BS $n$ assigns equal normalized resources to its associated users.

Else if BS $n$ is backhaul-constrained, similarly, there is
\begin{equation}
y_{nk}^*=\frac{1}{\nu_n^*R_{nk}W_n}=\frac{Z_{n,\text{bh}}}{|\mathcal{K}_n|R_{nk}W_n},\forall k\in\mathcal{K}_n, \label{t28}
\end{equation}
which means that BS $n$ averages its backhaul capacity among the associated users, and all of them achieve equal long-term service rate.

Moreover, in the extreme case that BS $n$ is both energy-constrained and backhaul-constrained, there are $\mu_n^*\ge 0$, $\nu_n^*\ge 0$, and both (\ref{t27}) and (\ref{t28}) are satisfied. Thus, we have
\begin{equation}
R_{nk}=\frac{Z_{n,\text{bh}}}{Q_{n,\text{P}}W_n},\forall k\in\mathcal{K}_n, \label{v1}
\end{equation}
which implies that all the associated users of BS $n$ have the same data rate. In this case, BS $n$ assigns equal normalized resources to the associated users, and the associated users achieve equal long-term service rate.

In summary, by applying the complementary slackness property, we can efficiently obtain the optimal resource partition of BSs. Moreover, since \textbf{P4} is a convex problem, the local optimum is also the unique global optimum. Furthermore, we have the following proposition.

\begin{proposition}
\label{prop1} Based on the rates of associated users, BS $n$ can optimally decide its resource partition:
\begin{equation}
y_{nk}^*=
\begin{cases}
\frac{Q_{n,\text{P}}}{|\mathcal{K}_n|},\forall k\in\mathcal{K}_n, &\text{if\,\,$\sum_{k\in\mathcal{K}_n}R_{nk}\leq\frac{|\mathcal{K}_n|Z_{n,\text{bh}}}{W_n Q_{n,\text{P}}}$},\\
\frac{Z_{n,\text{bh}}}{|\mathcal{K}_n|R_{nk}W_n},\forall k\in\mathcal{K}_n,&\text{otherwise}.
\end{cases} \label{t29}
\end{equation}
\end{proposition}

\begin{IEEEproof}
Please refer to the Appendix \ref{proof of pro1} for a proof.
\end{IEEEproof}

That is to say, to maximize the network utility, if $\sum_{k\in\mathcal{K}_n}R_{nk}\leq\frac{|\mathcal{K}_n|Z_{n,\text{bh}}}{W_nQ_{n,\text{P}}}$, BS $n$ divides its normalized resources constrained by both overall RBs and available energy equally among the associated users; otherwise, BS $n$ will average its backhaul capacity, and all the associated users achieve equal long-term service rate.

For clarity, we define BSs' resource partition indicator $\bm\varpi$ as:
\begin{equation}
\varpi_n=
\begin{cases}
1,&\text{if\,\,$\sum_{k\in\mathcal{K}_n}R_{nk}\leq\frac{|\mathcal{K}_n|Z_{n,\text{bh}}}{W_n Q_{n,\text{P}}}$},\\
0,&\text{otherwise}.
\end{cases} \label{t30}
\end{equation}
which will be used in the subsection \ref{distri algo}.

\subsection{User Association Given Resource Partition of BSs}
\label{user asso}
By substituting (\ref{t27}) and (\ref{t28}) into \textbf{P3} respectively, we can obtain the corresponding solutions for user association. Here, taking the resource allocation in (\ref{t27}) for example, we show the solution in detail. Specifically, the corresponding user association problem is:
\begin{align}
  \textbf{P5}:\max\limits_{\bm{x}}\, &\sum_{k\in\mathcal{K}}\sum_{n\in\mathcal{N}}x_{nk}\log \left(\frac{Q_{n,\text{P}}R_{nk}W_n}{\sum_{k\in\mathcal{K}}x_{nk}}\right) \label{t31}\\
  \mbox{s.t.}\,\, &\sum_{n\in\mathcal{N}}x_{nk}=1,\forall k\in\mathcal{K} \tag{32a}\\
  &x_{nk}\in \{0,1\},\forall n\in\mathcal{N},\forall k\in\mathcal{K} \tag{32b}
\end{align}

\textbf{P5} is still combinatorial due to binary variables $x_{nk}$s. Towards its solution, firstly, a set of auxiliary variables $\{\phi_n=\sum_{k\in\mathcal{K}}x_{nk}\}$ are introduced; then, to develop a distributed algorithm, we adopt Lagrange dual decomposition method whereby a Lagrange multiplier $\bm\upsilon$ is introduced to relax the coupled constraints $\phi_n=\sum_{k\in\mathcal{K}}x_{nk}, \forall n\in\mathcal{N}$.

The dual problem can be expressed as:
\begin{equation}
\textbf{P6}:\bm G:\,\min\limits_{\bm\upsilon}\,G(\bm\upsilon)=H(\bm\upsilon)+I(\bm\upsilon), \label{t32}
\end{equation}
where
\begin{align}
H(\bm\upsilon)=&\max\limits_{\bm x}\,\sum_{n\in\mathcal{N}}\sum_{k\in\mathcal{K}}x_{nk}\left[\log (Q_{n,\text{P}}R_{nk}W_n)-\upsilon_n\right] \label{t33}\\
\mbox{s.t.}\,\,\,&\sum_{n\in\mathcal{N}}x_{nk}=1,\forall k\in\mathcal{K} \tag{34a}\\
&x_{nk}\in \{0,1\},\forall n\in\mathcal{N},\forall k\in\mathcal{K} \tag{34b}
\end{align}
\begin{equation}
I(\bm\upsilon)=\max\limits_{\phi_n\leq K}\sum_{n\in\mathcal{N}}\phi_n\left[\upsilon_n-\log (\phi_n)\right]. \label{t34}
\end{equation}

According to the direct observation, $H(\bm\upsilon)$ can be further simplified to $\max\limits_{n}\left[\log (Q_{n,\text{P}}R_{nk}W_n)-\upsilon_n\right]$ which means user $k$ chooses one BS to maximize $\left[\log (Q_{n,\text{P}}R_{nk}W_n)-\upsilon_n\right]$. It is indeed an algorithm at user $k$'s side, and the optimal user association can be written as
\begin{equation}
x_{nk}=
\begin{cases}
1&\text{if $n=n_k$}\\
0&\text{otherwise,}
\end{cases}\label{t35}
\end{equation}
where $n_k=\mathop{\arg\max}_n\left\{\log (Q_{n,\text{P}}R_{nk}W_n)-\upsilon_n\right\}$.

For $I(\bm\upsilon)$, the optimum load $\phi_n$ can be calculated for BS $n$ by applying KKT condition, and we have
\begin{equation}
\phi_n=\min\{\exp(\upsilon_n-1),K\}. \label{t36}
\end{equation}
This is indeed an algorithm at BS $n$'s side.

Note that the dual function $G(\bm\upsilon)$ is not differentiable, as $H(\bm\upsilon)$ is a piecewise linear function and not differentiable. Therefore, we cannot use the usual gradient methods; instead, we will solve the dual problem using subgradient method.

It is easy to verify that $u_n(\upsilon_n)=\phi_n-\sum_{k\in\mathcal{K}}x_{nk}$ is a subgradient of dual function $G(\bm\upsilon)$ at point $\upsilon_n$. Thus, by subgradient method, we obtain the following algorithm for Lagrange multiplier $\upsilon_n$:
\begin{equation}
\upsilon_n(t+1)=\left[\upsilon_n(t)-\eta_n(t)\left(\phi_n(t)-\sum_{k\in\mathcal{K}}x_{nk}(t)\right)\right]^+, \label{t37}
\end{equation}
where $\eta_n(\tau)$ is a positive scalar stepsize, and ``$+$" denotes the projection onto the set ${\Re}^+$ of non-negative real numbers.

Similarly, the optimal user association for resource allocation in (\ref{t28}) can be written as:
\begin{equation}
\widetilde{x_{nk}}=
\begin{cases}
1&\text{if $n=\widetilde{n_k}$}\\
0&\text{otherwise,}
\end{cases}\label{t38}
\end{equation}
where $\widetilde{n_k}=\mathop{\arg\max}_n\left\{\log (Z_{n,\text{bh}})-\upsilon_n\right\}$, $\upsilon_n$ can be updated in the same way as in (\ref{t37}), and $\phi_n$ has the same expression as in (\ref{t36}).

It is easy to see that the multiplier $\bm\upsilon$ can be regarded as a message between users and BSs, and be interpreted as the service cost of BSs decided by their available resources, energy and backhaul. Moreover, it tradeoffs supply and demand, $\sum_{k\in\mathcal{K}}x_{nk}(t)$ is deemed the serving demand of BS $n$ and $\phi_n$ the service provided by BS $n$. In fact, (\ref{t37}) meets the law of supply and demand, which means that the service cost will go up if the demand $\sum_{k\in\mathcal{K}}x_{nk}(t)$ exceeds the supply $\phi_n$ and vice versa. Therefore, some overloaded BSs will increase their service price such that fewer users are associated with them, while other under-loaded ones will decrease the service prices to attract more users.

\subsection{A Distributed Algorithm for Backhaul-Aware Joint User Association and Resource Allocation}
\label{distri algo}
\begin{figure}[htbp]
  \centering
  \includegraphics[width=13.0cm]{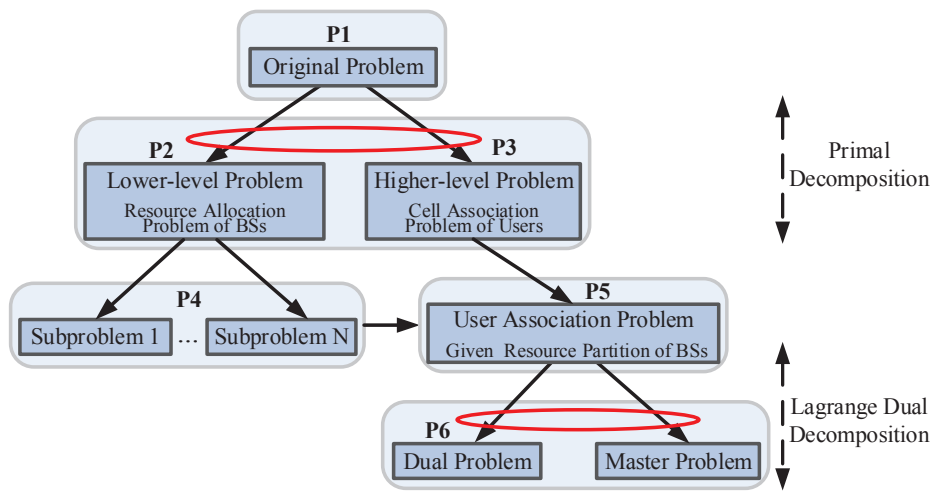}\\
  \caption{The problem decompositions} \label{fig:2}
\end{figure}

As presented in the above subsections, the formulated backhaul-aware joint user association and resource allocation problem is solved distributively through several decompositions. For clarity, the procedure of the problem decompositions is illustrated as in Fig. \ref{fig:2}. Specifically, the original problem \textbf{P1} is decomposed into a lower-level resource allocation problem \textbf{P2} and a higher-level user association problem \textbf{P3}. \textbf{P2} is further decomposed into subproblems \textbf{P4} for each BS. Solving \textbf{P4}, the optimal resource partition for a BS is obtained. Given solution to \textbf{P4}, which is a function of user association indicators, \textbf{P3} is transformed into \textbf{P5}. Then \textbf{P5} is solved through Lagrange dual decomposition, and its dual problem is \textbf{P6}. By solving \textbf{P6} with given solution to the master problem (optimal Lagrange multiplier), the optimal user association, together with the optimal resource partition of BSs, is determined.

Based on the above solution, we further present the distributed algorithm for backhaul-aware joint user association and resource allocation in energy-constrained HetNets, which can be divided into algorithms for each user and BS as in Algorithm \ref{algm1} and \ref{algm2}, respectively.

\begin{algorithm}
\caption{The Distributed Algorithm at User $k\in\mathcal{K}$}
\label{algm1}
\begin{algorithmic}
\IF{$t=0$}
\STATE Estimate $R_{nk}$, $\forall n\in\mathcal{N}$, by utilizing pilot signals.
\ELSE
\STATE Receive $\varpi_n(t)$ and $\upsilon_n(t)$, $\forall n\in\mathcal{N}$ broadcasted by all BSs and choose BS $n_k^*$ according to:
\begin{equation}
\begin{split}
n_k^*=&\mathop{\arg\max}_n\left\{\left[\log \left(R_{nk}Q_{n,\text{P}}W_n\right)-\upsilon_n(t)\right]\varpi_n(t)\right. \\
&\left.+\left[\log (Z_{n,\text{bh}})-\upsilon_n(t)\right]\left[1-\varpi_n(t)\right]\right\},\notag
\end{split}
\end{equation}

If there is more than one optimal association at the same time, a user can choose any one of them.
\STATE Feedback association information $x_{n_k^*k}(t)=1$ to BS $n_k^*$.
\ENDIF
\end{algorithmic}
\end{algorithm}

\begin{algorithm}
\caption{The Distributed Algorithm at BS $n\in\mathcal{N}$}
\label{algm2}
\begin{algorithmic}
\IF{$t=0$}
\STATE Initialize resource partition indicator $\varpi_n(0)$, stepsize $\eta_n(0)$ and $\upsilon_n(0)$.
\ELSE
\STATE Receive association information $x_{nk}(t)=1$, $\forall k\in\mathcal{K}$, determine the resource partition and update $\varpi_n(t+1)$ according to 
\begin{equation*}
\varpi_n(t+1)=
\begin{cases}
1,&\text{if\,\,$\sum_{k\in\mathcal{K}}x_{nk}(t)R_{nk}\leq\frac{|\mathcal{K}_n|Z_{n,\text{bh}}}{W_n Q_{n,\text{P}}}$},\\
0,&\text{otherwise}.
\end{cases}
\end{equation*}
\STATE Calculate $\phi_n(t)$ via
\begin{equation}
\phi_n(t)=\min\{\exp\left(\upsilon_n(t)-1\right),K\}.\notag
\end{equation}
\STATE Update Lagrange multiplier $\upsilon_n(t+1)$ via
\begin{equation}
\upsilon_n(t+1)=\left[\upsilon_n(t)-\eta_n(t)\left(\phi_n(t)-\sum_{k\in\mathcal{K}}x_{nk}(t)\right)\right]^+.\notag
\end{equation}
\STATE Broadcast the new $\varpi_n(t+1)$ and $\upsilon_n(t+1)$ to all users.
\ENDIF
\end{algorithmic}
\end{algorithm}

According to the above Algorithm \ref{algm1} and \ref{algm2}, in each iteration, each user reports its service request to only one BS to which it expects to connect, and each BS adjusts its service cost only relies on local information and then broadcasts it to all users. Therefore, the distributed method owns the amount of exchanged information of $\mathcal{O}(N+K)$ for each iteration, and the algorithms at a user's side and a BS's side have computation complexity of $\mathcal{O}(N)$ and $\mathcal{O}(K)$, respectively. Due to high convergence speed of the distributed algorithm, the iteration number is small. Consequently, the distributed algorithm may be more practicable for some cases, especially for large-scale problems, even if there exits more exchanged information.

After carrying out the algorithms at each user's side and each BS's side, the distributed algorithm can be guaranteed to converge, which will be proved by employing results of the subgradient method \cite{subg} in the following proposition.
\begin{proposition}
\label{prop2} Let $\bm\upsilon^*$ denote an optimal value of the dual variable. With constant stepsize, the proposed distributed algorithm is proved to converge statistically to $\bm\upsilon^*$, i.e., for any $\epsilon>0$, there exists a stepsize $\eta$, such that $\lim\sup_{t\to\infty}G\left(\overline{\bm\upsilon(t)}\right)-G(\bm\upsilon^*)\leq\epsilon$, where $\overline{\bm\upsilon(t)}=\frac{1}{t}\sum_{l=1}^t\upsilon(t)$; while for diminishing stepsize, the distributed algorithm is guaranteed to converge to the optimal value.
\end{proposition}

\begin{IEEEproof}
Please refer to the Appendix \ref{proof of pro2} for a proof.
\end{IEEEproof}

Therefore, based on results on convergence of the subgradient method, for constant stepsize, the proposed algorithm is guaranteed to converge to within a neighborhood of the optimal value; while for diminishing stepsize, the algorithm is guaranteed to converge to the optimal value. Moreover, since we focus on constant stepsize for theoretical proof, we will show the convergence with diminishing stepsize by numerical results in Section \ref{per eva}.

\section{Implementation Based on SDN}
\label{imple}
Although the proposed algorithm has provided a distributed user association and resource allocation solution, its application to the future wireless networks with an extreme BS deployment may still be inefficient due to the required interactive adjustments between users and BSs. In this section, based on the newly emerged SDN architecture, we design a vUARA scheme, which reduces the communication overhead over air interface and avoids the information leaking to users.

In the design of next-generation wireless networks, the challenges faced by current network architectures cannot be solved without a radical paradigm shift. Recently, by utilizing SDN, some architectures, such as software-define radio access network (SoftRAN) \cite{25} and SoftAir \cite{26}, have been proposed. Briefly, these architectures can accelerate the innovations for both hardware forwarding infrastructure and software networking algorithms through control and data separation, enable the efficient and adaptive sharing of network resources through network virtualization, achieve maximum spectrum efficiency through cloud-based collaborative baseband processing and enhance energy efficiency through the dynamic scaling of computing capacity of the software-defined BSs \cite{26}. Therefore, based on these architectures, we design a vUARA scheme. Mainly, by leveraging cloud computing and network virtualization, virtual users and vBSs can be generated in the radio access networks controller (RANC) to emulate a distributed joint user association and resource allocation solution that requires network measurements and iterative adjustments between users and BSs.

The implementation of vUARA is shown in Fig. \ref{fig:2}. Generally, it consists of three phases. The first phase is the initial user association and network measurements, during which the users and BSs measure and report their downlink SINRs, available energy and backhaul capacities to the RANC. In the second phase, the RANC firstly generates virtual users and vBSs, and then simulates the iterative user association and operation statuses (reflecting resource partition, load, and price of access) adjustments between users and BSs based on the information collected in the first phase. When the iterations converge, optimal cell association of users and resource allocation of BSs are derived. In the third phase, the RANC informs individual users and BSs about cell association and resource partition decisions. The major optimization of the vUARA scheme is done in the second phase. To be analytically tractable, we assume that (1) the RANC can successfully collect network information from all BSs and users, and (2) the data rates of users do not change within one user association process.
\vspace{-0.5cm}
\begin{figure}[htbp]
  \centering
  \includegraphics[width=11.5cm]{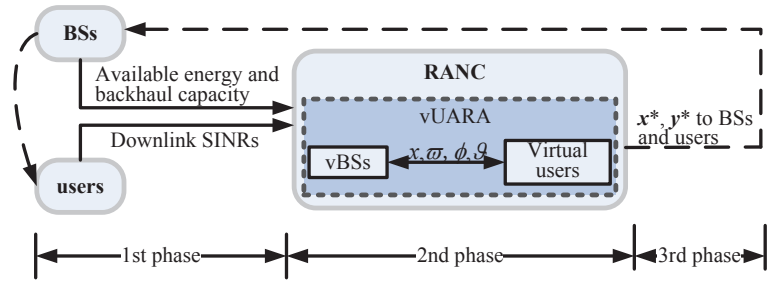}\\
  \caption{The implementation of vUARA} \label{fig:3}
\end{figure}

In summary, firstly, it is the RANC that jointly optimizes the user association and resource allocation based on network measurements. Secondly, instead of exchanging information over air interface that may introduce additional communication overhead and incur extra power consumption, the virtual users and vBSs can iteratively update their BS selections and operation statuses via a wired link, e.g., a message bus. Lastly, with vUARA, on one hand, the communication overhead over air interface is significantly reduced; on the other hand, the virtualization avoids leaking BSs' information to users since all the iterations are simulated in the RANC.

\section{Performance Evaluation}
\label{per eva}
\begin{figure}[htbp]
  \centering
  \includegraphics[width=12.0cm]{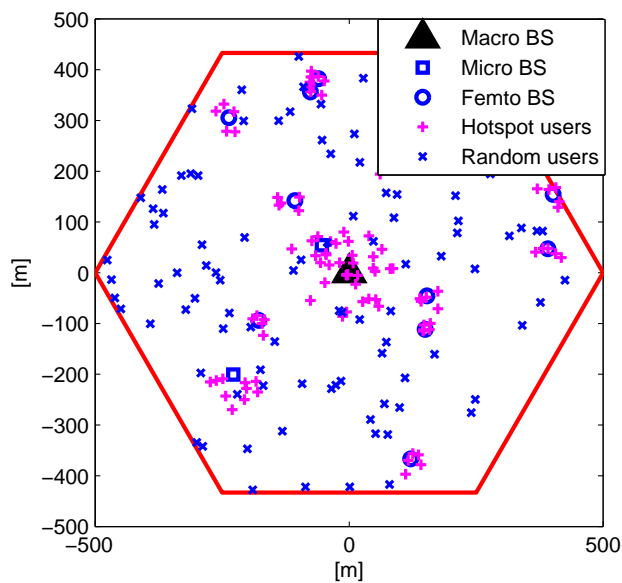}\\
  \caption{Simulation topology}\label{fig:4}
\end{figure}

In this section, we evaluate the performance of the proposed backhaul-aware joint user association and resource allocation algorithm for energy-constrained HetNets. Specifically, a 3-tier HetNet composed of one macro BS and several small cell (micro, femto) BSs is considered. The location of the macro BS is fixed and it forms a conventional cellular structure, while other small cell BSs are randomly located in an orthohexagonal area with side length $500m$. We assume that there exist two kinds of users in the scenario, i.e., hotspot users and random users. The former are located in the vicinity of each BS and the number of them is fixed, while the latter are randomly deployed in the whole area with varying number (denoted by $K_{\text{rand}}$). The simulation topology is given in Fig. \ref{fig:4}. Moreover, the energy and backhaul constraints of each BS and other basic parameters are presented in Table \ref{tab:1}. It is noted that, since the proposed algorithm deals with joint user association and resource allocation with energy constraints in an association period, we consider the overall energy constraint of a BS without its specific components.

\begin{table}
  \centering
  \caption{The basic parameters}
  \label{tab:1}
  \footnotesize
  \begin{tabular}{lccccccl}
  \toprule
  {\centering}Parameter    &macro BS &micro BS  &femto BS \\
  \midrule
  Number         &1       &4         &10  \\
  Hotspot users  &25     &10         &5  \\
  $P_{n0}$ (W)        &130     &56       &4.8 \\
  $\delta_n$   &4.7      &2.6       &8.0  \\
  Output power (dBm)	   &46    &35	    &20	  \\
  Path loss (dB)	 &\multicolumn{2}{c}{$34+40\log_{10}(d)$}  &\multicolumn{1}{c}{$37+30\log_{10}(d)$} \\
  $W_n$        &500      &100        &50 \\
  Available Energy (J/s) &300   &65    &5.6 \\
  $Z_{n,\text{bh}}$ (Mbps)  &2000    &200   &20 \\
  $\sigma^2$ (dBm) &\multicolumn{3}{c}{-111.45} \\
  \bottomrule
  \end{tabular}
\end{table}

\subsection{Convergence}
\begin{figure}[htbp]
  \centering
  \includegraphics[width=10.0cm]{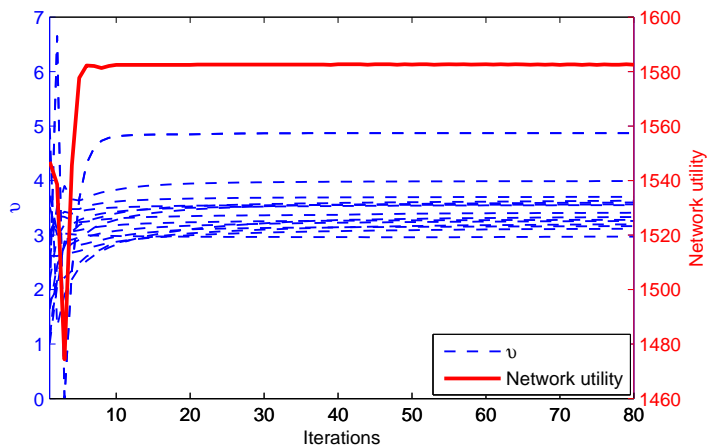}\\
  \caption{Lagrange multiplier $\bm\upsilon$ and network utility versus iterations ($K_{\text{rand}}=100$)}\label{fig:5}
\end{figure}

Given one random network realization, Fig. \ref{fig:5} presents the Lagrange multiplier $\bm\upsilon$ and network utility versus iterations. Specifically, the dashed lines refer to BSs' Lagrange multipliers with random initial values and distinct convergence process, and the solid line shows the network utility in terms of users' long-term service rate. Since the values of Lagrange multipliers and network utility are widely divergent, Fig. \ref{fig:5} adopts double ordinates, i.e., left ordinate and right ordinate, to refer to them respectively. The convergence of the proposed algorithm is demonstrated in Fig. \ref{fig:5}. Moreover, it is also shown that, the proposed algorithm may have a very fast convergence rate when parameters are set properly.

\subsection{Comparisons with Max-SINR and Range Expansion}
To demonstrate the superiority of the proposed algorithm, we compare it with the common user association schemes, i.e., max-SINR association, which associates users to the BS with the highest received SINR, and range expansion association, where the idea is to add a cell selection offset to the reference signals of the small cells in order to offload some traffic from the macrocells. Considering the energy and backhaul constraints, for max-SINR association and range expansion association, we adopt equal resource allocation among the users associated with the same BS and maximal achievable rate first (MARF) scheduling \cite{9} that selects users in descending order of achievable rates, given backhaul constraints. In addition, since the best offset (or biasing) factor is very difficult to obtain, here, we use offsets of 10dB and 12dB for micro BSs and femto BSs, respectively. All the following results are averaged over $10^4$ random network realizations.

\begin{figure}[htbp]
  \centering
  \includegraphics[width=10.0cm]{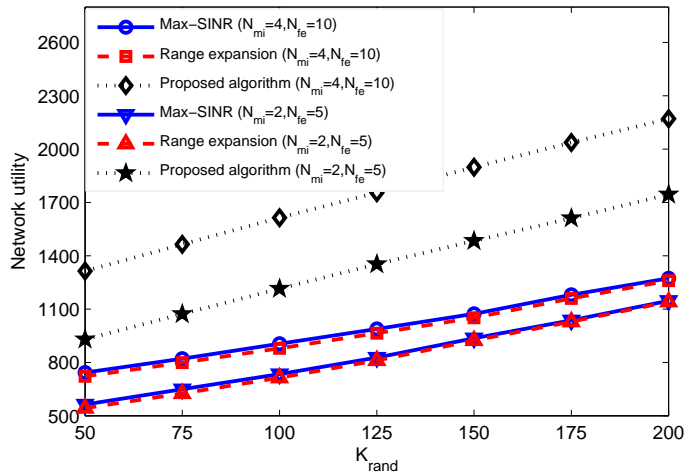}\\
  \caption{Network utility}\label{fig:6}
\end{figure}

\begin{figure}[htbp]
  \centering
  \includegraphics[width=10.0cm]{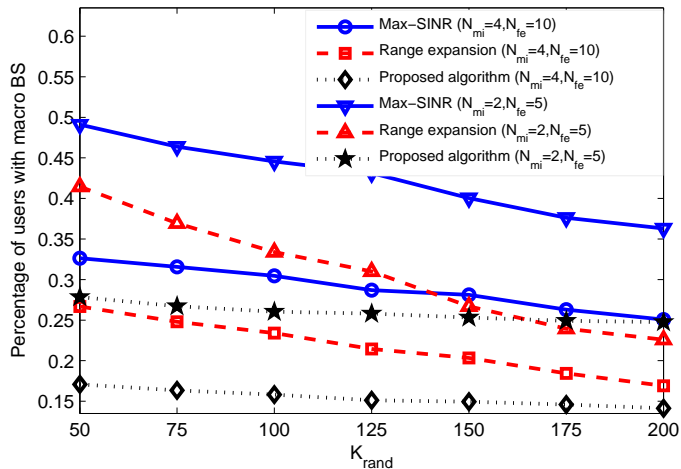}\\
  \caption{Percentage of users with macro BS}\label{fig:7}
\end{figure}

Given varying number of random users and small cell BSs, firstly, Figs. \ref{fig:6} and \ref{fig:7} respectively present the network utility and percentage of users associated with macro BS with the above three association methods. Specifically, Fig. \ref{fig:6} shows the significant improvement on network utility with the proposed algorithm, while the network utility with both max-SINR association and range expansion association is very low. Then, in Fig. \ref{fig:7}, it is easy to see that the percentage of users with the proposed algorithm is the lowest, while with the max-SINR association is highest. These facts confirm that (1) the max-SINR association will lead to very unbalanced load among BSs and greatly degraded network utility; (2) the range expansion association leads to better load balancing, but the improvement of load balancing may not overwhelm the degradation in SINR that certain users suffer; (3) the proposed algorithm gains significant improvement on both network utility and load balancing, since it takes both energy and backhaul constraints into consideration, and jointly optimizes the user association and resource partition. Moreover, it is obvious that, with fixed number of small cell BSs and moderate increase of the number of random users, the network utility increases given the above three association schemes, while the percentage of users associated with macro BS decreases. On the other hand, with more small cells deployed, all the association schemes achieve much better performance in terms of both network utility and load balancing.

\begin{figure}[htbp]
  \centering
  \includegraphics[width=10.0cm]{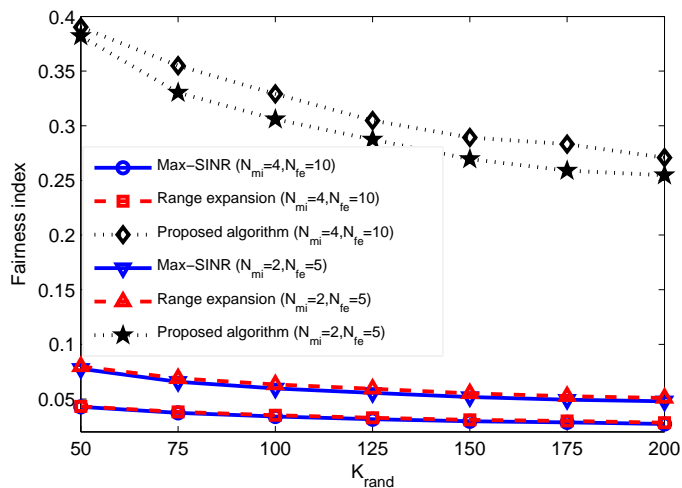}\\
  \caption{Fairness index among long-term service rates of users}\label{fig:8}
\end{figure}

Then, to measure the status of users' long-term service rates, we introduce the Jain's fairness index, that is, $\rho=(\sum_{k\in\mathcal{K}}R_k)^2/(K\sum_{k\in\mathcal{K}}R_k^2)$. The larger $\rho$ that belongs to the interval $[1/K,1]$ means more balanced long-term service rates among the users. Given the above three association schemes, Fig. \ref{fig:8} presents fairness index among long-term service rates of users. It is observed that, (1) the fairness index of the proposed algorithm is much greater than that of either max-SINR association or range expansion association; and (2) compared with max-SINR, the fairness index of range expansion association is only a little higher. These are due to the facts that, on one hand, the proposed algorithm jointly optimizes the user association and resource allocation under constraints of resources, energy and backhaul; on the other hand, the offset factor set here may be far from the ``best", which degrades the performance of the range expansion association. Moreover, it is worth mentioning that, for max-SINR association and range expansion association with equal resource allocation and MARF scheduling, some users may be dropped since their BSs are running out of backhaul; while with the proposed algorithm, all the users are accommodated. In addition, both the number of users and that of small cell BSs have impacts on user fairness, that is, for the proposed algorithm, both the increased deployment of small cells and decreased access of users improve the user fairness. While for max-SINR association and range expansion association, the increase of the number of small cells will not improve the user fairness index, which further proves that these two association schemes have little help in providing user fairness.

\subsection{Effects of Energy and Backhaul Constraints}
\begin{figure}[htbp]
  \centering
  \includegraphics[width=10.0cm]{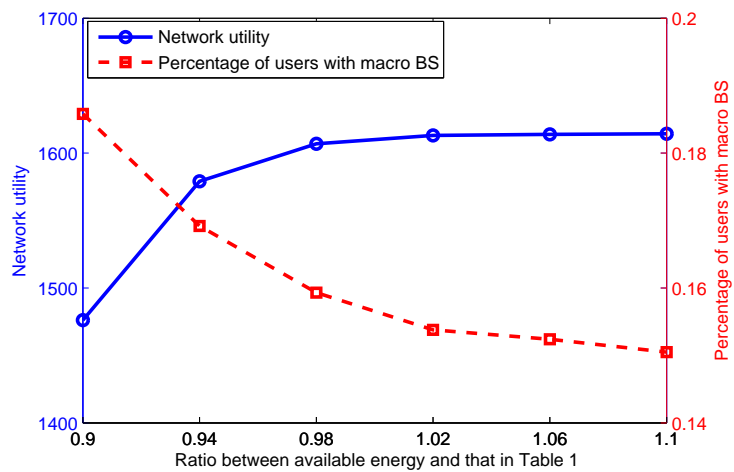}\\
  \caption{Effects of energy constraints ($K_{\text{rand}}=100$)}\label{fig:9}
\end{figure}

\begin{figure}[htbp]
  \centering
  \includegraphics[width=10.0cm]{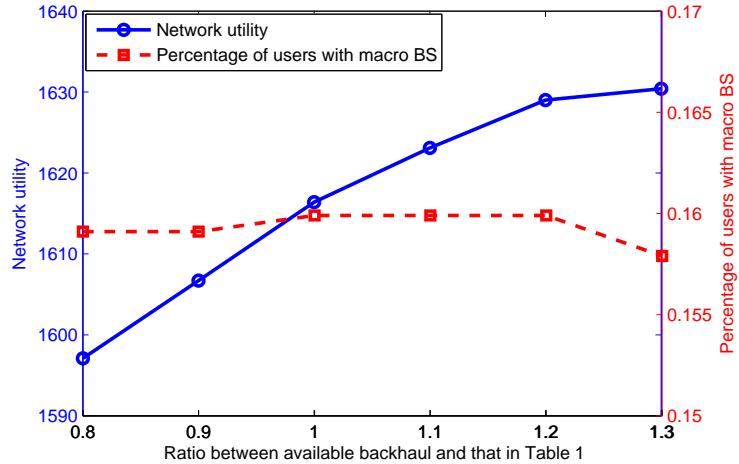}\\
  \caption{Effects of backhaul constraints ($K_{\text{rand}}=100$)}\label{fig:10}
\end{figure}

Since constraints of both energy and backhaul are taken into consideration, we illustrate their effects on algorithm performance. All the results are averaged over $10^4$ random network realizations. Firstly, Fig. \ref{fig:9} shows the network utility and percentage of users with macro BS versus ratio between the available energy and that in Table \ref{tab:1}. We can see that, with the increase of available energy at each BS, the network utility firstly increases then stays almost unchanged due to backhaul constraints. While the percentage of users associated with macro BS firstly decreases since more users will associate with the small cell BSs once they have more or enough energy. However, because of the backhaul constraints, the percentage may not keep on decreasing. Then, the network utility and percentage of users with macro BS versus ratio between the backhaul capacities and those in Table \ref{tab:1} are given in Fig. \ref{fig:10}. It is observed that, with the increase of backhaul capacity at each BS, firstly, the network utility increases rapidly, while the percentage of users with macro BS remains almost unchanged; then the network utility has slower increase, while the percentage decreases. This is due to the fact that, at the beginning, all the BSs may be greatly constrained by their backhaul, thus, with the increase of backhaul capacities, the network utility increases fast, while the percentage of users with macro BS may not decrease; then, when backhaul capacities are large enough, available energy becomes the constraints for all the BSs, especially for macro BS, as a result, the network utility may not keep on increasing, while the percentage of users with macro BS decreases. However, Figs. \ref{fig:9} and \ref{fig:10} only show the rough tendency with the increase of available energy and backhaul capacities, respectively, and the results greatly depend on the parameter settings given in Table {\ref{tab:1}}.

\begin{figure}[htbp]
  \centering
  \includegraphics[width=10.0cm]{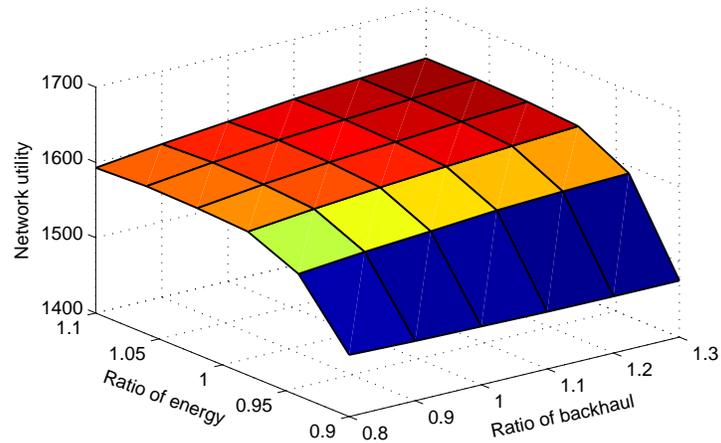}\\
  \caption{Effects of energy and backhaul constraints on network utility ($K_{\text{rand}}=100$)}\label{fig:11}
\end{figure}

\begin{figure}[htbp]
  \centering
  \includegraphics[width=10.0cm]{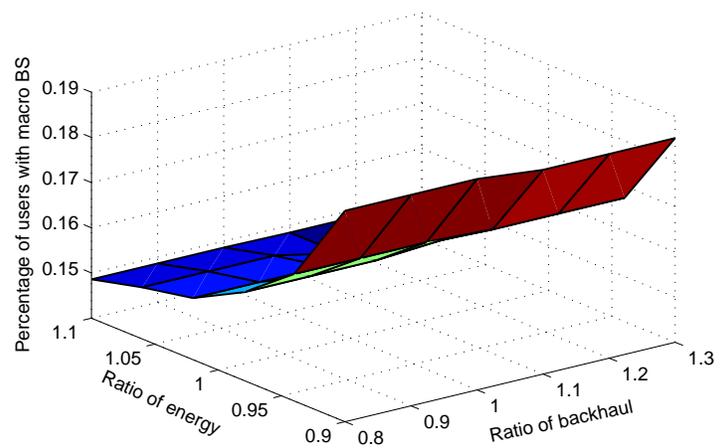}\\
  \caption{Effects of energy and backhaul constraints on percentage of users with macro BS ($K_{\text{rand}}=100$)}\label{fig:12}
\end{figure}

Moreover, in Figs. \ref{fig:11} and \ref{fig:12}, the network utility and percentage of users with macro BS are respectively presented with varying constraints of energy and backhaul. Particularly, the change of energy and backhaul constraints is the same as that in Figs. \ref{fig:9} and \ref{fig:10}, respectively. It is observed that, with fixed backhaul (energy), the network utility and percentage of users with macro BS have the same tendency as those in Fig. \ref{fig:9} (Fig. \ref{fig:10}). Besides, Figs. \ref{fig:11} and \ref{fig:12} also show that, compared to the network utility improvement with increasing backhaul capacities of all the BSs, the network utility increases rapidly with the increase of available energy; and the percentage of users with macro BS reduces much faster with increasing available energy of all the BSs, compared to its change with the increase of backhaul capacities. These may due to the fact that, the available energy of BSs affects the algorithm performance together with the available resources of BSs. Thus, much greater impacts are observed.
Nevertheless, the constraints of both backhaul and energy affect the algorithm performance, and the impacts relate to the parameter settings given in Table \ref{tab:1}. In a word, the specific effects of energy and backhaul constraints are coupled and complicated; the network design should take both of them into account.
\vspace{-0.1cm}

\section{Conclusions}
\label{conclu}
In this paper, given a HetNet powered by hybrid energy, we focus on the backhaul-aware joint user association and resource allocation problem. To balance network-wide performance and user fairness, we formulate an online network utility maximization problem reflecting PF, which has tightly coupled variables (both binary and continuous) in the constraints of resources, energy and backhaul. Then, by adopting some decomposition methods, the condition for two kinds of resource partition of a BS is efficiently obtained, and a completely distributed algorithm is developed. Finally, the convergence of the distributed algorithm is proved by employing results of the subgradient method. Moreover, based on SDN architecture, we develop a vUARA scheme for future wireless networks with an extremely dense BS deployment. Lastly, numerical results indicate that, compared with max-SINR and range expansion, the proposed algorithm provides a significant improvement on network utility, load balancing and user fairness.
\appendices

\section{Proof of Proposition 1}
\label{proof of pro1}
If BS $n$ is energy-constrained, according to (\ref{t27}) and the second inequality of (\ref{t25}), we have
\begin{equation}
\sum_{k\in\mathcal{K}_n}\frac{Q_{n,\text{P}}}{|\mathcal{K}_n|}R_{nk}W_n\leq Z_{n,\text{bh}}, \label{t39}
\end{equation}
i.e.,
\begin{equation}
\sum_{k\in\mathcal{K}_n}R_{nk}\leq\frac{|\mathcal{K}_n|Z_{n,\text{bh}}}{W_nQ_{n,\text{P}}}, \label{t40}
\end{equation}
which means that, to maximize the network utility, if $\sum_{k\in\mathcal{K}_n}R_{nk}\leq\frac{|\mathcal{K}_n|Z_{n,\text{bh}}}{W_nQ_{n,\text{P}}}$, BS $n$ will divide its available resources equally among the associated users; otherwise, BS $n$ is backhaul-constrained, and according to (\ref{t28}), it will average the backhaul capacity and provide all the associated users with equal long-term service rate.

\section{Proof of Proposition 2}
\label{proof of pro2}
Given constant stepsize $\eta_n(t)=\eta$, according to the updating of $\bm\upsilon$, we have
\begin{equation}
\begin{split}
&\left\|\bm\upsilon(t+1)-\bm\upsilon^*\right\|^2_2\\
&=\left\|\left[\bm\upsilon(t)-\eta\bm u(t)\right]^+-\bm\upsilon^*\right\|^2_2\\
&\leq \left\|\bm\upsilon(t)-\eta\bm u(t)-\bm\upsilon^*\right\|^2_2\\
&=\left\|\bm\upsilon(t)-\bm\upsilon^*\right\|^2_2-2\eta\bm u(t)^{\text{T}}\left(\bm\upsilon(t)-\bm\upsilon^*\!\right)\!+\!\eta^2\left\|\bm u(t)\right\|^2_2\\
&\leq\!\left\|\bm\upsilon(t)\!-\!\bm\upsilon^*\right\|^2_2\!-\!2\eta\left(\!G(\bm\upsilon(t))-G(\bm\upsilon^*)\right)+\eta^2\left\|\bm u(t)\right\|^2_2,
\end{split} \label{t41}
\end{equation}
where the last inequality follows from the definition of subgradient. Applying the inequalities recursively, we obtain
\begin{equation}
\begin{split}
&2\sum_{l=1}^t\eta\left(G(\bm\upsilon(l))\!-\!G(\bm\upsilon^*)\right)\\
&\!\leq\!-\!\left\|\bm\upsilon(t+1)\!-\!\bm\upsilon^*\right\|^2_2\!+\!\left \|\bm\upsilon(1)\!-\!\bm\upsilon^*\right\|^2_2\!+\!\sum_{l=1}^t\eta^2\left\|\bm u(l)\right\|^2_2\\
&\!\leq\!\left\|\bm\upsilon(1)\!-\!\bm\upsilon^*\right\|^2_2\!+\!\sum_{l=1}^t\eta^2\left\|\bm u(l)\right\|^2_2.
\end{split} \label{t42}
\end{equation}

From this inequality, we have
\begin{equation}
\frac{1}{t}\sum_{l=1}^t\left(G(\bm\upsilon(l))\!-\!G(\bm\upsilon^*)\right)\leq\frac{\!\left\|\bm\upsilon(1)-\bm \upsilon^*\right\|^2_2}{2t\eta}+\frac{\!\sum_{l=1}^t\!\eta\left\|\bm u(l)\right\|^2_2}{2t}. \label{t43}
\end{equation}

Since $G(\bm\upsilon)$ is a convex function, by Jensen's inequality, there is
\begin{equation}
G\left(\overline{\bm\upsilon(t)}\right)-G(\bm\upsilon^*)\leq\frac{\left\|\bm\upsilon(1)-\bm \upsilon^*\right\|^2_2}{2t\eta}+\frac{\sum_{l=1}^t\eta\left\|\bm u(l)\right\|^2_2}{2t}. \label{t44}
\end{equation}

When both $\phi_n(l)$ and $\sum_{k=1}^Kx_{nk}(l)$ are bounded, $\left\|\bm u(l)\right\|^2_2$ is bounded too, i.e., $\sup\limits_l\left\|\bm u(l)\right\|^2_2\leq c$ , where $c$ is a scalar. Then,
\begin{equation}
G\left(\overline{\bm\upsilon(t)}\right)-G(\bm\upsilon^*)\leq\frac{\left\|\bm\upsilon(1)-\bm\upsilon^*\right\|^2_2}{2t\eta}+\frac{\eta c}{2}. \label{t45}
\end{equation}

Therefore, $\lim\sup_{t\to\infty}G\left(\overline{\bm\upsilon(t)}\right)-G(\bm\upsilon^*)\leq\epsilon$, where $\epsilon=\eta c/2$. That is to say, given constant stepsize, the algorithm converges statistically to within $\eta c/2$ of the optimal value. Besides, if stepsize $\eta$ is small enough, the algorithm converges to the optimal value.
\end{spacing}
\begin{spacing}{1.5}

\end{spacing}

\begin{thebibliography}{99}
\bibitem{1}
J. G. Andrews, S. Buzzi, W. Choi, S. Hanly, A. Lozano, A. C. Soong, and J. C. Zhang, ``What will 5G be?'' \emph{IEEE J. Sel. Areas Commun.}, vol. 32, no. 6, pp. 1065--1082, June 2014.
\bibitem{vf1}
Z. Su, Q, Xu, Q. Qi, ``Big data in mobile social networks: a QoE-oriented framework,'' \emph{IEEE Network}, vol. 30, no. 1, pp. 52-57, Jan.-Feb. 2016.
\bibitem{2}
J. G. Andrews, ``Seven ways that HetNets are a cellular paradigm shift,'' \emph{IEEE Commun. Mag.}, vol. 51, no. 3, pp. 136--144, Mar. 2013.
\bibitem{3}
A. Damnjanovic, J. Montojo, Y. Wei, T. Ji, T. Luo, M. Vajapeyam, T. Yoo, O. Song, and D. Malladi, ``A survey on 3GPP heterogeneous
networks,'' \emph{IEEE Trans. Wireless Commun.}, vol. 18, no. 3, pp. 10--21, June 2011.
\bibitem{4}
E. Hossain, M. Rasti, H. Tabassum, and A. Abdelnasser, ``Evolution toward 5G multi-tier cellular wireless networks: An interference management perspective
,'' \emph{IEEE Wireless Commun.}, vol. 21, no. 3, pp. 1536--1284, June 2014.
\bibitem{vf2}
Y. Zou, J. Lam, Y.Niu, D. Li, ``Constrained predictive control synthesis for quantized systems with Markovian data loss,'' \emph{Automatica}, vol. 55,  pp. 217--225, May 2015.
\bibitem{v1} 
G. Iosifidis, L. Gao, J. Huang, and L. Tassiulas, ``An iterative double auction for mobile data offloading,'' in \emph{Proc. Modeling \& Optimization in Mobile, Ad Hoc \& Wireless Networks}, May 2013, pp. 154--161.
\bibitem{v2}
S. He, Y. Huang, S. Jin, and L. Yang, ``Coordinated beamforming for energy efficient transmission in multicell multiuser systems,¡¯¡¯ \emph{IEEE Trans. Commun.}, vol. 61, no.12, pp. 4961--4971, Dec. 2013.
\bibitem{v3}
H. Farhadi, C. Wang, and M. Skoglund, ``Distributed transceiver design and power control for wireless MIMO interference networks,¡¯¡¯ \emph{IEEE Trans. Commun.}, vol. 14, no. 3, pp. 1199--1212, Mar. 2015.
\bibitem{vf3}
Y. Song, C. Zhang, and Y. Fang, ``Minimum energy scheduling in multi-hop wireless networks with retransmissions,'' \emph{IEEE Trans. Wireless Commun.}, vol. 9, no.1, pp. 348--355, Jan. 2010.
\bibitem{5} 
Z. Hasan, H. Boostanimehr, and V. K. Bhargava, ``Green cellular networks: A survey, some research issues and challenges,'' \emph{IEEE Commun. Surveys and Tuts.}, vol. 13, no. 4, pp. 524--540, Fourth Quarter 2011.
\bibitem{6}
H. S. Dhillon, Y. Li, P. Nuggehalli, Z. Pi, and J. G. Andrews, ``Fundamentals of heterogeneous cellular networks with energy harvesting,'' \emph{IEEE Trans. Wireless Commun.}, vol. 13, no. 5, pp. 2782--2797, May 2014.
\bibitem{vf4}
S. He, J. Chen, F. Jiang, D. K. Yau, G. Xing, and Y. Sun, ``Energy provisioning in wireless rechargeable sensor networks,'' \emph{IEEE Trans. Mobile Computing}, vol. 12, no. 10, pp. 1931--1942, Oct. 2013.
\bibitem{7} 
G. Piro, M. Miozzo, G. Forte, N. Baldo, L. Grieco, G. Boggia, and P. Dini, ``HetNets powered by renewable energy sources: Sustainable next-generation cellular networks,'' \emph{IEEE Internet Computing}, vol. 17, no. 1, pp. 32--39, Jan.-Feb. 2013.
\bibitem{8}
J. Andrews, S. Singh, Q. Ye, X. Lin, and H. Dhillon, ``An overview of load balancing in HetNets: Old myths and open problems,'' \emph{IEEE Trans. Wireless Commun.}, vol. 21, no. 2, pp. 18--25, Apr. 2014.
\bibitem{9}
T. Zhou, Y. Huang, and L. Yang, ``QoS-aware user association for load balancing in heterogeneous cellular networks,'' in \emph{Proc. IEEE VTC Fall}, Sep. 2014, pp. 1--5.
\bibitem{v4}
H. Dai, Y. Huang, and L.Yang, ``Game theoretic max-logit learning approaches for joint base station selection and resource allocation in heterogeneous networks,¡¯¡¯ \emph{IEEE J. Sel. Areas Commun.}, vol. 33, no. 6, pp. 1068--1081, May 2015.
\bibitem{10} 
S. Ulukus, A. Yener, E. Erkip, O. Simeone, M. Zorzi, P. Grover, and K. Huang, ``Energy harvesting wireless communications: A review of recent advances,'' \emph{IEEE J. Sel. Areas Commun.}, vol. 33, no. 3, pp. 360--381, Mar. 2015.
\bibitem{11}
S. Hur, T. Kim, D. J. Love, J. V. Krogmeier, T. A. Thomas, and A. Ghosh, ``Millimeter wave beamforming for wireless backhaul and access in small cell networks,'' \emph{IEEE Trans. Commun.}, vol. 61, no. 10, pp. 4391--4403, Oct. 2013.
\bibitem{12}
X. Ge, H. Cheng, M. Guizani, and T. Han, ``5G wireless backhaul networks: Challenges and research advances,'' \emph{IEEE Networks}, vol. 28, no. 6, pp. 6--11, Nov.-Dec. 2014.
\bibitem{13}
K. Shen and W. Yu, ``Distributed pricing-based user association for downlink heterogeneous cellular networks,'' \emph{IEEE J. Sel. Areas Commun.}, vol. 32, no. 6, pp. 1100--1113, June 2014.
\bibitem{14}
H. Boostanimehr and V. K. Bhargava, ``Unified and distributed QoS-driven cell association algorithms in heterogeneous networks,'' \emph{IEEE Trans. Wireless Commun.}, vol. 14, no. 3, pp. 1650--1662, Jan. 2015.
\bibitem{15}
N. Wang, E. Hossain, and V. K. Bhargava, ``Joint downlink cell association and bandwidth allocation for wireless backhauling in two-tier HetNets with large-scale antenna arrays,'' http://arxiv.org/abs/1501.00078, 2014.
\bibitem{16}
C. Kim, R. Ford, Y. Qi, and S. Rangan, ``Joint interference and user association optimization in cellular wireless networks,'' \emph{2014 48th Asilomar Conference on Signals, Systems and Computers}, pp. 511--515, Nov. 2014.
\bibitem{17}
C. Zhe and R. Adve, ``Joint user association and resource allocation in small cell networks with backhaul constraints,'' in \emph{Proc. CISS}, Mar. 2014, pp. 1--6.
\bibitem{18} 
J. Rubio, A. Pascual Iserte, J. Del Olmo, and J. Vidal, ``User association for load balancing in heterogeneous networks powered with energy harvesting sources,'' in \emph{Proc. IEEE GLOBECOM}, Dec. 2014, pp. 1227--1232.
\bibitem{vf5}
Q. Han, B. Yang, C. Chen, and X. Guan, ``Energy-aware and QoS-aware load balancing for HetNets powered by renewable energy,'' \emph{Computer Networks}, vol. 94, pp. 250--262, Jan. 2016.
\bibitem{19}
T. Han and N. Ansari, ``Green-energy aware and latency aware user associations in heterogeneous cellular networks,'' in \emph{Proc. IEEE GLOBECOM}, Dec. 2013, pp. 4946--4951.
\bibitem{20}
T. Han and N. Ansari, ``On optimizing green energy utilization for cellular networks with hybrid energy supplies,'' \emph{IEEE Trans.
Wireless Commun.}, vol. 12, no. 8, pp. 3872--3882, Aug. 2013.
\bibitem{21}
T. Han and N. Ansari, ``A traffic load balancing framework for software-defined radio access networks powered by hybrid energy sources,'' \emph{IEEE/ACM Trans. Networking}, vol. PP, no. 99, pp. 1--14, Mar. 2015.
\bibitem{22}
T. Han and N. Ansari, ``Network utility aware traffic loading balancing in backhaul-constrained cache-enabled small cell networks with hybrid power supplies,'' http://arxiv.org/abs/1409.8267, 2014.
\bibitem{23}
G. Auer, V. Giannini, C. Desset, I. Godor, P. Skillermark, M. Olsson, M. A. Imran, D. Sabella, M. J. Gonzalez, O. Blume, and A. Fehske,
``How much energy is needed to run a wireless network?'' \emph{IEEE Trans. Wireless Commun.}, vol. 18, no. 5, pp. 40--49, Oct. 2011.
\bibitem{24}
J. Gong, S. Zhou, and Z. Niu, ``Optimal power allocation for energy harvesting and power grid coexisting wireless communication systems,'' \emph{IEEE Trans.
Wireless Commun.}, vol. 61, no. 7, pp. 3040--3049, Jul. 2013.
\bibitem{minlp}
M. Hasan and E. Hossain, ``Distributed resource allocation for relay-aided device-to-device communication: A message passing approach,'' \emph{IEEE Trans.
Wireless Commun.}, vol. 13, no. 11, pp. 6326--6341, Nov. 2014.
\bibitem{subg}
S. Boyd and A. Mutapcic, ``Subgradient methods,'' \emph{Lecture notes of EE364b, Stanford University}, Winter Quarter, 2007.
\bibitem{25}
A. Gudipati, D. Perry, L. E. Li, and S. Katti, ``SoftRAN: Software defined radio access network,'' in \emph{Proc. 2nd ACM SIGCOMM Workshop on Hot Topics in Software Defined Networking}, Hong Kong, 2013, pp. 25--30.
\bibitem{26}
I. F. Akyildiz, P. Wang, and S. C. Lin, ``SoftAir: A software defined networking architecture for 5G wireless systems,'' \emph{Computer Networks}, vol. 85, pp. 1--18, July 2015.
\end{thebibliography}
\end{document}